\documentclass[twocolumn]{aastex631} %
\usepackage{amsmath}
\usepackage{graphicx}
\usepackage{graphbox}
\usepackage{natbib}
\usepackage{booktabs}
\usepackage{xspace}
\usepackage{array}
\usepackage{multirow}

\newcommand{\oj}{OJ\,287\xspace}

\def\smili{\texttt{SMILI}\xspace}
\def\ehtim{\texttt{eht-imaging}\xspace}
\def\difmap{\texttt{DIFMAP}\xspace}

\received{21 March, 2022}
\accepted{27 April, 2022}
\submitjournal{ApJ}

\shorttitle{The first GMVA+ALMA observations of \oj}
\shortauthors{Zhao et al.}

\graphicspath{{./}{figures/}}

\begin{document}

\title{Unravelling the Innermost Jet Structure of \oj with the First GMVA+ALMA Observations}

\author[0000-0002-4417-1659]{Guang-Yao Zhao}
\affiliation{Instituto de Astrof\'{\i}sica de Andaluc\'{\i}a-CSIC, Glorieta de la Astronom\'{\i}a s/n, E-18008 Granada, Spain}  

\author[0000-0003-4190-7613]{Jos\'e L. G\'omez}
\affiliation{Instituto de Astrof\'{\i}sica de Andaluc\'{\i}a-CSIC, Glorieta de la Astronom\'{\i}a s/n, E-18008 Granada, Spain}

\author[0000-0002-8773-4933]{Antonio Fuentes}
\affiliation{Instituto de Astrof\'{\i}sica de Andaluc\'{\i}a-CSIC, Glorieta de la Astronom\'{\i}a s/n, E-18008 Granada, Spain}

\author[0000-0002-4892-9586]{Thomas P. Krichbaum}
\affiliation{Max-Planck-Institut f\"ur Radioastronomie, Auf dem H\"ugel 69, D-53121 Bonn, Germany}

\author[0000-0002-1209-6500]{Efthalia Traianou}
\affiliation{Instituto de Astrof\'{\i}sica de Andaluc\'{\i}a-CSIC, Glorieta de la Astronom\'{\i}a s/n, E-18008 Granada, Spain}

\author[0000-0001-7361-2460]{Rocco Lico}
\affiliation{Instituto de Astrof\'{\i}sica de Andaluc\'{\i}a-CSIC, Glorieta 
de la Astronom\'{\i}a s/n, E-18008 Granada, Spain}
\affiliation{INAF-Istituto di Radioastronomia, Via P. Gobetti 101, I-40129 Bologna, Italy}

\author[0000-0001-6083-7521]{Ilje Cho}
\affiliation{Instituto de Astrof\'{\i}sica de Andaluc\'{\i}a-CSIC, Glorieta de la Astronom\'{\i}a s/n, E-18008 Granada, Spain}

\author[0000-0001-9503-4892]{Eduardo Ros}
\affiliation{Max-Planck-Institut f\"ur Radioastronomie, Auf dem H\"ugel 69, D-53121 Bonn, Germany}

\author{S. Komossa}
\affiliation{Max-Planck-Institut f\"ur Radioastronomie, Auf dem H\"ugel 69, D-53121 Bonn, Germany}


\author[0000-0002-9475-4254]{Kazunori Akiyama}
\affiliation{Massachusetts Institute of Technology Haystack Observatory, 99 Millstone Road, Westford, MA 01886, USA}
\affiliation{National Astronomical Observatory of Japan, 2-21-1 Osawa, Mitaka, Tokyo 181-8588, Japan}
\affiliation{Black Hole Initiative at Harvard University, 20 Garden Street, Cambridge, MA 02138, USA}

\author[0000-0001-6988-8763]{Keiichi Asada}
\affiliation{Institute of Astronomy and Astrophysics, Academia Sinica, 11F of 
Astronomy-Mathematics Building, AS/NTU No. 1, Sec. 4, Roosevelt Rd, Taipei 10617, Taiwan, R.O.C.}

\author[0000-0002-9030-642X]{Lindy Blackburn}
\affiliation{Black Hole Initiative at Harvard University, 20 Garden Street, Cambridge, MA 02138, USA}
\affiliation{Center for Astrophysics $|$ Harvard \& Smithsonian, 60 Garden Street, Cambridge, MA 02138, USA}

\author[0000-0001-9240-6734]{Silke Britzen}
\affiliation{Max-Planck-Institut f\"ur Radioastronomie, Auf dem H\"ugel 69, D-53121 Bonn, Germany}

\author[0000-0002-5182-6289]{Gabriele Bruni}
\affiliation{INAF -- Istituto di Astrofisica e Planetologia Spaziali, via Fosso del Cavaliere 100, 00133 Roma, Italy}

\author[0000-0002-2079-3189]{Geoffrey B. Crew}
\affiliation{Massachusetts Institute of Technology Haystack Observatory, 99 Millstone Road, Westford, MA 01886, USA}

\author[0000-0001-6982-9034]{Rohan Dahale}
\affiliation{Instituto de Astrof\'{\i}sica de Andaluc\'{\i}a-CSIC, 
Glorieta de la Astronom\'{\i}a s/n, E-18008 Granada, Spain}
\affiliation{Indian Institute of Science Education and Research Kolkata, Mohanpur, Nadia, West Bengal 741246, India}

\author{Lankeswar Dey}
\affiliation{Department of Astronomy and Astrophysics, Tata Institute of Fundamental Research, Mumbai 400005, India}

\author[0000-0003-2492-1966]{Roman Gold}
\affiliation{CP3-Origins, University of Southern Denmark, Campusvej 55, DK-5230 Odense M, Denmark}

\author{Achamveedu Gopakumar}
\affiliation{Department of Astronomy and Astrophysics, Tata Institute of Fundamental Research, Mumbai 400005, India}

\author[0000-0002-5297-921X]{Sara Issaoun}
\affiliation{Center for Astrophysics $|$ Harvard \& Smithsonian, 60 Garden Street, Cambridge, MA 02138, USA}
\affiliation{Department of Astrophysics, Institute for Mathematics, Astrophysics and Particle
Physics (IMAPP), Radboud University, P.O. Box 9010, 6500 GL Nijmegen, The Netherlands}
\affiliation{NASA Hubble Fellowship Program, Einstein Fellow}

\author[0000-0001-8685-6544]{Michael Janssen}
\affiliation{Max-Planck-Institut f\"ur Radioastronomie, Auf dem H\"ugel 69, D-53121 Bonn, Germany}

\author[0000-0001-6158-1708]{Svetlana Jorstad}
\affiliation{Institute for Astrophysical Research, Boston University, 725 Commonwealth Ave., Boston, MA 02215, USA}
\affiliation{Astronomical Institute, St. Petersburg University, Universitetskij pr., 28, Petrodvorets,198504 St.Petersburg, Russia}

\author[0000-0001-8229-7183]{Jae-Young Kim}
\affiliation{Department of Astronomy and Atmospheric Sciences, Kyungpook National University, Daegu 702-701, Republic of Korea}
\affiliation{Korea Astronomy and Space Science Institute, Daedeok-daero 776, Yuseong-gu, Daejeon 34055, Republic of Korea}
\affiliation{Max-Planck-Institut f\"ur Radioastronomie, Auf dem H\"ugel 69, D-53121 Bonn, Germany}

\author[0000-0002-7029-6658]{Jun Yi Koay}
\affiliation{Institute of Astronomy and Astrophysics, Academia Sinica, 11F of Astronomy-Mathematics Building, AS/NTU No. 1, Sec. 4, Roosevelt Rd, Taipei 10617, Taiwan, R.O.C.}

\author[0000-0001-9303-3263]{Yuri Y. Kovalev}
\affiliation{Lebedev Physical Institute of the Russian Academy of Sciences, Leninsky prospekt 53, 119991 Moscow, Russia}
\affiliation{Moscow Institute of Physics and Technology,
  Institutsky per. 9, Dolgoprudny, Moscow region, 141700, Russia}
\affiliation{Max-Planck-Institut f\"ur Radioastronomie, Auf dem H\"ugel 69, D-53121 Bonn, Germany}

\author[0000-0002-3723-3372]{Shoko Koyama}
\affiliation{Niigata University, 8050 Ikarashi-nino-cho, Nishi-ku, Niigata 950-2181, Japan}
\affiliation{Institute of Astronomy and Astrophysics, Academia Sinica, 11F of
Astronomy-Mathematics Building, AS/NTU No. 1, Sec. 4, Roosevelt Rd, Taipei 10617, 
Taiwan, R.O.C.}

\author[0000-0003-1622-1484]{Andrei P. Lobanov}
\affiliation{Max-Planck-Institut f\"ur Radioastronomie, Auf dem H\"ugel 69, D-53121 Bonn, Germany}

\author[0000-0002-5635-3345]{Laurent Loinard}
\affiliation{Instituto de Radioastronom\'{\i}a y Astrof\'{\i}sica, Universidad Nacional Aut\'onoma de M\'exico, Morelia 58089, M\'exico}
\affiliation{Instituto de Astronom\'{\i}a, Universidad Nacional Aut\'onoma de M\'exico, CdMx 04510, M\'exico}

\author[0000-0002-7692-7967]{Ru-Sen Lu}  
\affiliation{Shanghai Astronomical Observatory, Chinese Academy of Sciences, 80 Nandan Road, Shanghai 200030, People's Republic of China}
\affiliation{Key Laboratory of Radio Astronomy, Chinese Academy of Sciences, Nanjing 210008,
People’s Republic of China}
\affiliation{Max-Planck-Institut f\"ur Radioastronomie, Auf dem H\"ugel 69, D-53121 Bonn, Germany}

\author[0000-0001-9564-0876]{Sera Markoff}
\affiliation{Anton Pannekoek Institute for Astronomy, University of Amsterdam, Science Park 904, 1098 XH, Amsterdam, The Netherlands}
\affiliation{Gravitation and Astroparticle Physics Amsterdam (GRAPPA) Institute, University of Amsterdam, Science Park 904, 1098 XH Amsterdam, The Netherlands}

\author[0000-0001-7396-3332]{Alan P. Marscher}
\affiliation{Institute for Astrophysical Research, Boston University, 725 Commonwealth Ave., Boston, MA 02215, USA}

\author[0000-0003-3708-9611]{Iv\'an Martí-Vidal}
\affiliation{Departament d'Astronomia i Astrof\'{\i}sica, Universitat de Val\`encia, C. Dr. Moliner 50, E-46100 Burjassot, Val\`encia, Spain}
\affiliation{Observatori Astronòmic, Universitat de Val\`encia, C. Catedr\'atico Jos\'e Beltr\'an 2, E-46980 Paterna, Val\`encia, Spain}

\author[0000-0002-8131-6730]{Yosuke Mizuno}
\affiliation{Tsung-Dao Lee Institute, Shanghai Jiao Tong University, Shengrong Road 520, Shanghai, 201210, People’s Republic of China}
\affiliation{School of Physics and Astronomy, Shanghai Jiao Tong University, 800 Dongchuan Road, Shanghai, 200240, People’s Republic of China}
\affiliation{Institut f\"ur Theoretische Physik, Goethe-Universit\"at Frankfurt, Max-von-Laue-Stra{\ss}e 1, D-60438 Frankfurt am Main, Germany}

\author[0000-0001-6558-9053]{Jongho Park}
\affiliation{Institute of Astronomy and Astrophysics, Academia Sinica, 11F of 
Astronomy-Mathematics Building, AS/NTU No. 1, Sec. 4, Roosevelt Rd, Taipei 10617, Taiwan, R.O.C.}

\author[0000-0001-6214-1085]{Tuomas Savolainen}
\affiliation{Aalto University Department of Electronics and Nanoengineering, PL 15500, FI-00076 Aalto, Finland}
\affiliation{Aalto University Mets\"ahovi Radio Observatory, Mets\"ahovintie 114, FI-02540 Kylm\"al\"a, Finland}
\affiliation{Max-Planck-Institut f\"ur Radioastronomie, Auf dem H\"ugel 69, D-53121 Bonn, Germany}

\author[0000-0003-3658-7862]{Teresa Toscano}
\affiliation{Instituto de Astrof\'{\i}sica de Andaluc\'{\i}a-CSIC, Glorieta de la Astronom\'{\i}a s/n, E-18008 Granada, Spain}


\begin{abstract}
  \noindent
 We present the first very-long-baseline interferometric (VLBI) observations of the blazar \oj carried out jointly with the Global Millimeter VLBI Array (GMVA) and the phased Atacama Large Millimeter/submillimeter Array (ALMA) at 3.5\,mm on April 2, 2017. 
  Participation of phased-ALMA not only has improved the GMVA north-south resolution by a factor of $\sim$,3, but also has enabled fringe detection with signal-to-noise ratios up to 300 at baselines longer than 2\,G$\lambda$. 
  The high sensitivity has motivated us to image the data with the newly developed regularized maximum likelihood imaging methods, revealing the innermost jet structure with unprecedentedly high angular resolution.
  Our images reveal a compact and twisted jet extending along the northwest direction with two 
  bends   
  within the inner 200\,{\textmu}as that resembles a precessing jet in projection. The component at the southeastern end shows a compact morphology and high brightness temperature, and is identified as the VLBI core. An extended jet feature that lies at $\sim$\,200\,{\textmu}as northwest of the core shows a conical shape in both total and linearly polarized intensity, and a bimodal distribution of the linear polarization electric vector position angle.
  We discuss the nature of this feature by comparing our observations with models and simulations of oblique and recollimation shocks with various magnetic field configurations. 
  Our high-fidelity images also enabled us to search for possible jet features from the secondary supermassive black hole (SMBH) and test the SMBH binary hypothesis proposed for this source.

\end{abstract}




\section{Introduction}
\label{Sec:Intro}

The BL\,Lac type object \oj \citep[$z=0.306$;][]{1989A&AS...80..103S} is a well-studied low synchrotron peaked BL\,Lac object (LBL) that has attracted great interest as it shows quasi-periodic optical outbursts with a cycle of about 12 years.
These outbursts appear to come in pairs with separations of one to two years and have been suggested to originate due to the presence of a supermassive binary black hole (SMBBH) system at its center~\citep[e.g.,][]{1988ApJ...325..628S,1996ApJ...460..207L}. 
According to this model, the observed quasi-periodic double-peaked optical outbursts are triggered when the secondary supermassive black hole (SMBH) impacts the accretion disk of the primary in its orbit.  
Further advances of this model have accounted for general relativistic effects and the parameters are also further constrained with follow-up observations~\citep[e.g.,][]{2008Natur.452..851V, 2011ApJ...729...33V, 2018ApJ...866...11D}. 
The model requires a compact binary with a major axis of the orbit of 0.112\,pc \citep[corresponding to an angular scale of $\sim$\,26\,{\textmu}as; e.g.,][]{2008Natur.452..851V}, featuring a very massive primary BH of $1.8 \times 10^{10} M_\odot$, and a secondary of $1.5 \times 10^{8} M_\odot$~\citep[e.g.,][]{2012MNRAS.427...77V, 2018ApJ...866...11D}.
This model is not only successful in reproducing the observed light curves of \oj, but also in predicting impact outbursts that were later confirmed by observations~\citep[e.g.,][]{2006ApJ...643L...9V, 2016ApJ...819L..37V, 2020ApJ...894L...1L, 2020MNRAS.498L..35K}.
Independent of the binary model of \oj, dedicated multi-wavelength observation and modeling of the \oj (MOMO) project has led to the discovery of several bright flare events and long-lasting deep fades, and monitoring spectroscopy of the last two decades has established \oj as one of the most spectrally variable blazars in the soft X-ray band~\citep[e.g.,][]{2017IAUS..324..168K, 2021ApJ...923...51K, 2021Univ....7..261K, 2021MNRAS.504.5575K}.

Another observational signature of \oj is that the position angle (PA) of the parsec-scale jet was found to be ``wobbling" by previous very long baseline interferometric (VLBI) observations \citep[e.g.,][]{Tateyama_2004,2012ApJ...747...63A, galaxies5010012, 2018MNRAS.478.3199B}.   
Such changes of the inner jet PA could also be explained by the SMBBH model~\citep[e.g.,][]{2021MNRAS.503.4400D}, but alternative models could not be fully ruled out.
For instance, \citet{2012ApJ...747...63A} suggest instabilities coupled to the accretion disk as likely origin for the non-periodic changes in the inner jet orientation. \citet{2018MNRAS.478.3199B} suggest the flux variation could be explained by viewing angle changes and Doppler beaming effects of a precessing jet. The precession could be driven by either the binary motion~\citep[e.g.,][]{2021MNRAS.503.4400D} or the Lense-Thirring effect due to the misalignment between the BH spin and the accretion disc~\citep[e.g.,][]{2020MNRAS.499..362C,2018MNRAS.474L..81L}.

The massive central black hole (BH), the relatively low redshift, and the bright close to line-of-sight relativistic jet also make \oj one of the nearest high-luminosity AGN in which the magnetic launching and acceleration of jets can be studied through high-resolution VLBI observations.
Two competing scenarios have been proposed for the formation of relativistic jets. 
The main difference between them is whether the magnetic fields are twisted by the rotational energy of the BH \citep[BZ model;][]{1977MNRAS.179..433B} or its accretion disk \citep[BP model;][]{1982MNRAS.199..883B}.  
It is also possible that both mechanisms are at work~\citep[e.g.,][]{2000A&A...358..104C}.
In the innermost region of the jet, the plasma flow is accelerated and collimated in the presence of a spiral magnetic field, while the jet expands in width and propagates downstream into the interstellar space. 
The disruption of the accretion flow and the interaction with the ambient medium often result in the formation of moving and standing shocks.
The detailed process of jet formation, acceleration, and collimation remains unclear as it requires extremely high angular resolution to probe into the innermost region in the vicinity of the central black hole.

High-resolution VLBI observations are ideal for probing the compact structure near the central engine. 
Previous VLBI observations of \oj have provided key information on the parsec-scale structure and dynamics of the jet~\citep[e.g.][]{2017A&A...597A..80H,galaxies5010012, 2018MNRAS.478.3199B}. 
In particular, \citet{2022ApJ...924..122G} recently presented 22\,GHz images of \oj with unprecedented angular resolution for the source obtained with the \textsl{RadioAstron} space-ground VLBI observations. 
The images revealed a progressive bending of the inner jet with increasing angular resolution by comparison with multi-band ground-based VLBI images. 
The inner jet components show high brightness temperatures that exceed the inverse Compton limit, indicating strong Doppler boosting in the jet. The polarized images show electric vector position angles (EVPAs) aligned with the jet axis, which indicates the jet has a predominantly toroidal magnetic field. Multi-frequency analysis shows hints for a rotation measure gradient across the jet, which suggests the VLBI core is threaded by a helical magnetic field.

VLBI observations at wavelengths shorter than 7\,mm hold the potential of probing areas closer to the central engine that are optically thick at lower frequencies \citep[see e.g.,][]{2017A&ARv..25....4B}. Previous VLBI observations at 3.5\,mm with the Global Millimeter VLBI Array (GMVA) show the existence of quasi-stationary components and changes in the morphology and PA in the innermost jet region \citep[e.g.,][]{2017A&A...597A..80H}.
However, most of the previous GMVA observations are limited in sensitivity due to typically shorter atmospheric coherence times, lower antenna efficiencies, and thus higher system equivalent flux densities (SEFDs) compared to longer wavelengths.
Participation of large sensitive stations in mm-VLBI observations are desirable alongside with further developments of the instruments and calibration methods~\citep[e.g.,][]{Rioja:2011fz, 2017Galax...5....9R, 2018AJ....155...26Z}. 

In this paper, we present the first VLBI observations of \oj with the GMVA and phased Atacama Large Millimeter/submillimeter Array (ALMA) on April 2, 2017. These observations are accompanied by a multi-wavelength campaign including the first 1.3\,mm observation of the source with the Event Horizon Telescope~\citep[EHT;][]{2019ApJ...875L...2E}, the results of which will be presented in a forthcoming paper. The campaign was carried out during a major outburst event of \oj in 2016-17 with the largest X-ray outburst recorded so far~\citep[][]{2017IAUS..324..168K, 2021MNRAS.504.5575K} and the first very high energy (VHE) flare detection~\citep{2017ATel10051....1M}.

We summarize the details of the GMVA+ALMA observations and the methods we use to calibrate, image, and analyze the data in \autoref{Sec:Obs};
we present our observational results including total intensity and linear polarization images in \autoref{Sec:Results};
in \autoref{Sec:Discussion}, we discuss the nature of the components in the jet and possible constraints on the theoretical models, followed by a summary in \autoref{Sec:Summary}.


\section{Observations and data analysis}
\label{Sec:Obs}
  
\begin{figure}[t]
\centering
\includegraphics[width=0.99\columnwidth]{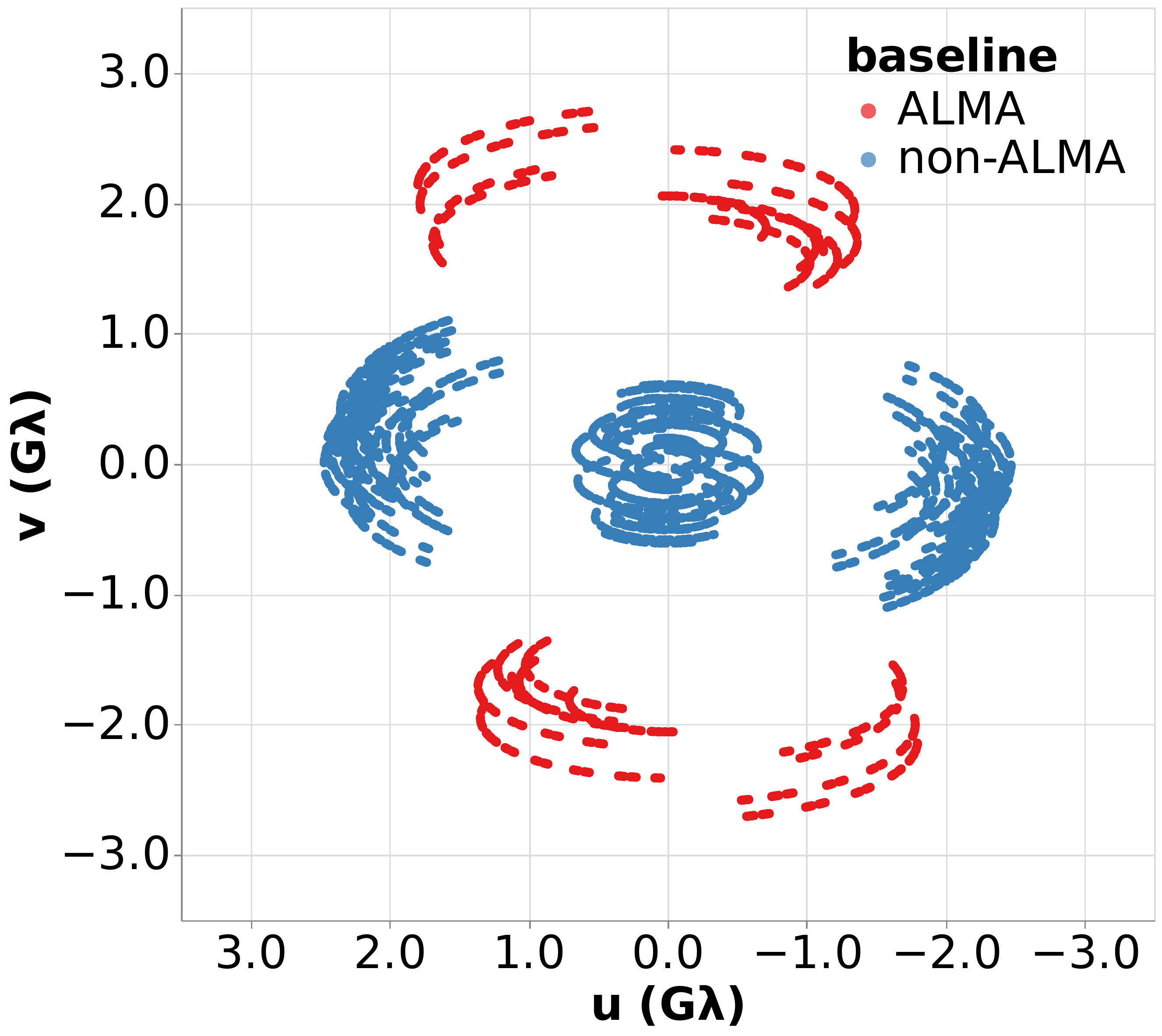} \hfill
\caption
  { 
  $(u,v)$-coverage of the fringe-fitted interferometric visibilities of \oj, observed with GMVA+ALMA on April 2, 2017 at 86\,GHz. The baselines to ALMA are plotted in red color and the other GMVA baselines are plotted in blue.  
  }
\label{fig:uvplot}
\end{figure}
  
  
\begin{figure}[t]
\centering
\includegraphics[width=0.999\linewidth]{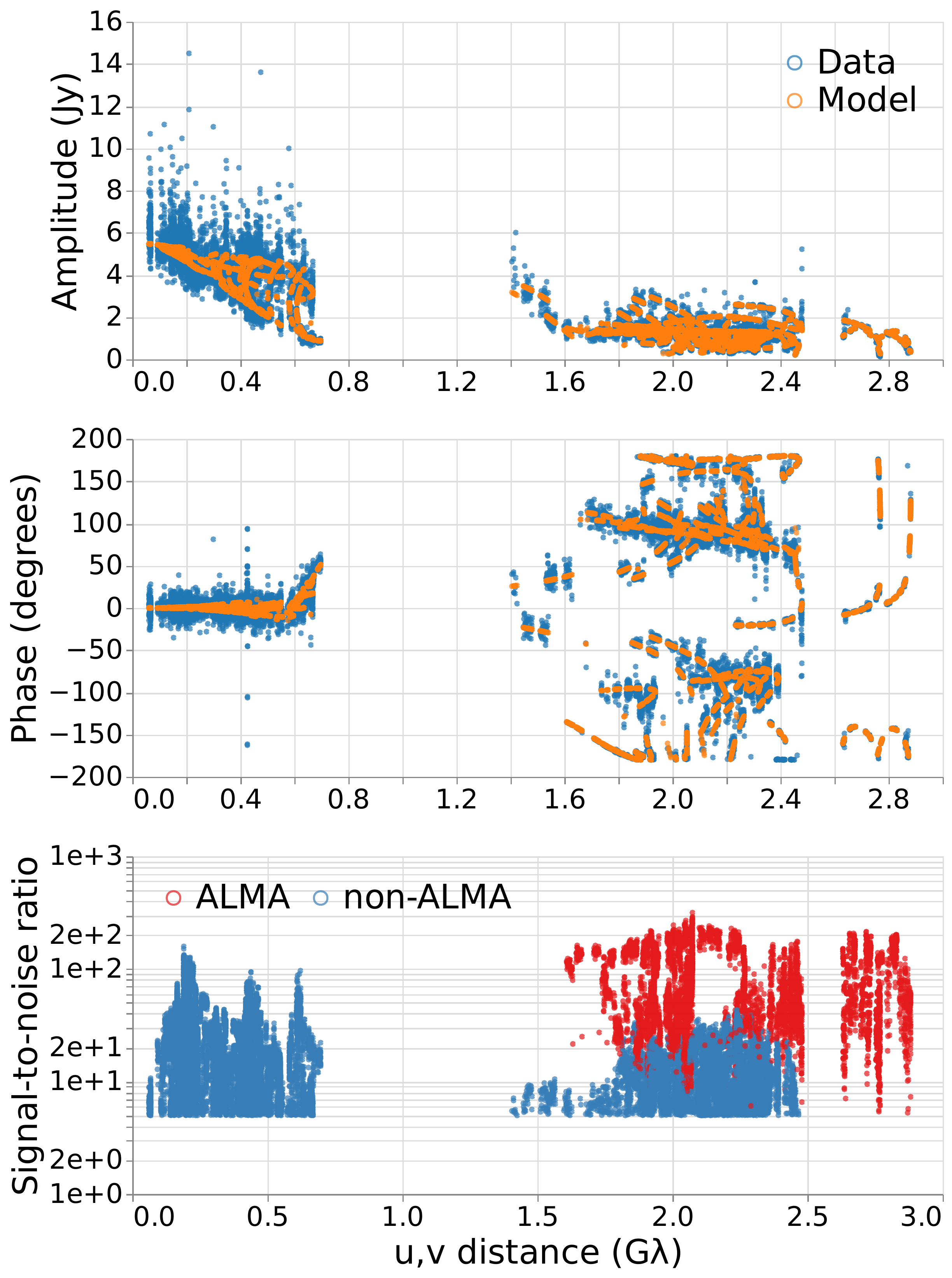}\hfill
\caption
  { 
  Self-calibrated visibility amplitudes (top) and phases (middle) as a function of $(u,v)$-distance of the GMVA+ALMA observation of \oj on April 2, 2017 at 86\,GHz. 
  The data were averaged every 15 seconds and all channels in each IF are averaged.
  Over-plotted in orange are the fit to the data of the reconstructed image obtained with \smili.
  The bottom panel shows the fringe signal-to-noise ratio as a function of $(u,v)$-distance, with the data on ALMA baselines plotted in red and the other baselines in blue.
  }
\label{fig:radplot}
\end{figure}
  
In this section, we describe the details of our 3.5\,mm observations of \oj with GMVA + ALMA, the data calibration procedure, and the methods used to obtain subparsec-scale images of \oj.

\subsection{Observations}
We carried out high-resolution VLBI observations towards \oj at 3.5\,mm with GMVA on April 2, 2017. These observations mark the first VLBI observations with the phased-ALMA which consists of 37 ALMA antennas and is equivalent to a 70-meter dish~\citep[][]{2019ApJ...875L...2E}. The participating stations also include 8 Very Long Baseline Array (VLBA) stations and 5 European stations (Effelsberg, IRAM-30m, Mets\"ahovi, Onsala, Yebes-40m). 
The on-source time was around 375 minutes between UT 17 and UT 7 the next day (April 3).
 
Most stations had good or typical weather conditions during the observation except for the VLBA Mauna Kea (MK) and Pie Town (PT) stations, which resulted in few fringe detections with limited signal-to-noise ratios (S/N) on baselines to these two stations. No fringes were found on baselines to Mets\"ahovi due to a faulty backend setup. All data were recorded in full polarization mode, with most stations recorded on a circular polarization basis, while the ALMA data were converted from a mixed linear-circular basis to circular polarization mode using \texttt{PolConvert}~\citep[][]{2016A&A...587A.143M}. Yebes-40m telescope recorded only left hand circular polarization (LCP). The bandwidth and frequency range recorded are not the same at all stations\footnote{The recorded bandwidth for each station is as follows: ALMA 32$\times$62.5MHz, VLBA 2$\times$128MHz, most European stations 1$\times$512MHz}.   
Only the common frequency ranges among all participating stations are used in later processing.
\subsection{Data Reduction}
Data correlation was performed with the DiFX correlator \citep{Deller_2007} at the Max-Planck-Institut f\"ur Radioastronomie in Bonn, Germany. %
The final correlated data have a total bandwidth of 232\,MHz which were further divided into four 58\,MHz intermediate frequency (IF) bands. 
 
The post-correlation dataset was then processed with the \texttt{ParselTongue} \citep{2006ASPC..351..497K} {$\mathcal{AIPS}$} \citep{2003ASSL..285..109G} interface for fringe-fitting and a priori amplitude calibration. 
We first performed parallactic angle correction with the {$\mathcal{AIPS}$} task, \textsc{clcor}~\footnote{We note the mount types for IRAM-30m (Nasmyth-Left) and Yebes-40m (Nasmyth-Right) are different from the rest of antennas in the array (altitude-azimuth). The Yebes-40m data were not used for polarimetric analysis as they were only recorded in LCP.}, and manual phase calibration using short segments of data to remove instrumental phase offset between different IFs. 
We then perform a global fringe-fitting of the data using the task \textsc{fring}  with a solution interval of 10 seconds and sub-intervals down to 2 seconds and by integrating over the whole 232\,MHz bandwidth and averaging parallel-hand polarizations (RR \& LL). 

 The $(u,v)$-coverage towards \oj for all baselines with fringe detections is shown in \autoref{fig:uvplot}.   
 We note that the participation of ALMA has provided an increase in the north-south resolution by a factor of $\sim$\,3 for observations of \oj.
 ALMA has also significantly improved fringe detection due to its high sensitivity (see \autoref{fig:radplot}) with the maximum fringe S/N reaching $\sim$\,350 at baselines longer than 1.5\,G$\lambda$.  
  
A priori amplitude calibration was performed in {$\mathcal{AIPS}$} with the task \textsc{apcal} 
by multiplying the system temperatures ($T_{\text{sys}}$) and gain curves of each antenna. 
Opacity corrections were applied to stations that measure the system temperatures with the noise diode method (VLBA \& Effelsberg). 
For ALMA, IRAM-30m, and Yebes-40m, the $T_{\text{sys}}$ measurements were performed using the hot / cold method and therefore already included the opacity correction. The ALMA $T_{\text{sys}}$ values have also taken into account the phasing efficiencies derived during the quality assurance and \texttt{PolConvert} processes~\citep[e.g.,][]{2019PASP..131g5003G}.
The cross-hand phase and delay offsets of the reference station were calibrated using the {$\mathcal{AIPS}$} procedure, \textsc{vlbacpol}.

After the {$\mathcal{AIPS}$} calibration, the data were averaged in time (with an interval of 15\,s) and frequency (with all channels within each IF averaged) for further processing.
  
\subsection{Imaging $\&$ Model-fitting}
We performed imaging and self-calibration of the data independently with three different imaging softwares: \difmap, \ehtim, and \smili. 
\difmap is the software commonly used for the conventional CLEAN method for interferometric imaging~\citep{1995BAAS...27..903S}. It interactively establishes a collection of point source models from the inverse Fourier transform of the visibilities, i.e., the dirty map. 
CLEAN windows, which define the regions to search for CLEAN components, are used during our imaging process. 
Phase-only self-calibration is performed after each step of cleaning. 
Amplitude and phase self-calibration is performed once a good fit to the visibilities is established through the multiple steps of cleaning and phase self-calibration.  
We repeat the clean and self-calibration loops several times during our imaging process by gradually decreasing the solution interval of the amplitude and phase self-calibration.
On the other hand, the regularized maximum likelihood (RML) methods, employed by the \ehtim~\citep{Chael2016,2018ApJ...857...23C} and \smili~\citep{2017AJ....153..159A} libraries, reconstruct images by minimizing an objective function which is a weighted combination of $\chi^2$ of the data and various regularizer terms. 
The data terms may include the closure quantities \citep[closure phases and amplitudes; e.g.,][]{2017isra.book.....T}, visibility amplitudes, and complex visibilities.
Common regularizers include the maximum entropy~\citep[e.g.,][]{2018ApJ...857...23C}, the $\ell _1$-norm~\citep[e.g.,][]{Honma:2014jv, 2017AJ....153..159A}, the total variation (TV) and the total squared variation (TSV) of the brightness~\citep[e.g.,][]{2018ApJ...858...56K}. 
With RML methods, it is possible to achieve an angular resolution a few times finer than the nominal interferometric beam~\citep[e.g.,][]{2017AJ....153..159A, 2019ApJ...875L...4E}. 
During our imaging process with \ehtim and \smili, we started with a Gaussian prior image and reconstruct images with only the closure quantities, or a combination of  closure quantities and low-weighted visibility amplitudes. After a few iterations of imaging and self-calibrating, we include full complex visibilities into the optimization process, further constraining the reconstructed images.
To determine the best set of regularizer combinations, we survey a range of different weights of each regularizer, in total $\sim$\,128 combinations, and select the one that results in the best fit to the closure quantities. 

After imaging of the total intensity, we estimate the instrumental polarimetric leakage (known as D-terms) for each station using the self-calibrated data of \oj. This process was carried out independently with two pipelines: the {$\mathcal{AIPS}$} task \textsc{lpcal}, and the \ehtim library, each based on a particular set of self-calibrated dataset generated during the total intensity imaging process, i.e., \difmap and \ehtim, respectively.
Both approaches provide consistent values of D-terms. Details of leakage calibration are described in \autoref{appendix::dterms}.
Polarization imaging of the \textsc{lpcal} processed data was carried out with \difmap. 
With \ehtim, the imaging were performed iteratively with the D-term calculation.
Calibration of the absolute orientation of the EVPAs was performed through comparison with the ALMA array data \citep{Goddi21}.

We also carried out non-imaging analysis of the data to measure the properties of the jet.
We perform circular Gaussian model-fitting to the \smili self-calibrated visibility data with \difmap.
The results indicate that the jet structure can be represented by four Gaussian components. 
We label the components following the convention described in \citet[][]{2022ApJ...924..122G}.
The total flux, size, and position offset with respect to the core (the component at the southeastern end of the jet; see \autoref{Sec:Results} below) of all components are listed in \autoref{tab:modelfit}. 
The uncertainties of the fitted parameters are derived following the equations outlined in \citet{2019A&A...622A..92N}.
%

\begin{deluxetable*}{cccccc}
\tablecaption{Model-fitting parameters of \oj with GMVA+ALMA on April 2, 2017. \label{tab:modelfit}}
\setlength{\tabcolsep}{2mm}
\tablewidth{0pt}
\tablehead{
\colhead{Comp} &  \colhead{S} & \colhead{r} & \colhead{PA} & \colhead{FWHM} & \colhead{$T^\mathrm{obs}_{b}$} \\
\colhead{Name} & \colhead{(Jy)} & \colhead{({\textmu}as)} & \colhead{($^{\circ}$)} & \colhead{({\textmu}as)} & \colhead{($10^{10}$ K)} \\
\colhead{(1)} & \colhead{(2)} & \colhead{(3)} & \colhead{(4)} & \colhead{(5)} & \colhead{(6)}  
}
\startdata
C0  & 1.25 $\pm$ 0.13 &   0.0           &    0.0           & 31.1 $\pm$ 2.4 & 21.2 $\pm$ 4.1 \\
C1  & 0.96 $\pm$ 0.10 &  63.9 $\pm$ 2.4 &  -41.5 $\pm$ 2.2 & 28.9 $\pm$ 2.4 & 18.8 $\pm$ 3.9 \\
C2a & 2.49 $\pm$ 0.25 & 172.9 $\pm$ 2.4 &  -40.0 $\pm$ 0.8 & 44.4 $\pm$ 2.5 & 20.6 $\pm$ 3.1 \\
C2b & 0.51 $\pm$ 0.05 & 212.7 $\pm$ 3.5 &  -47.7 $\pm$ 0.9 & 42.5 $\pm$ 4.6 &  4.7 $\pm$ 1.2 \\
\hline
\enddata
\tablecomments{Columns from left to right: (1) Component ID, (2) Flux density, (3) Radial distance from the core component (C0), (4) Position angle, (5)Component full width at half maximum (FWHM), (6) Observed brightness temperature 
}
\end{deluxetable*}

\section{Results}
\label{Sec:Results}

\begin{figure*}[t]
\centering
\includegraphics[width=1.0\textwidth]{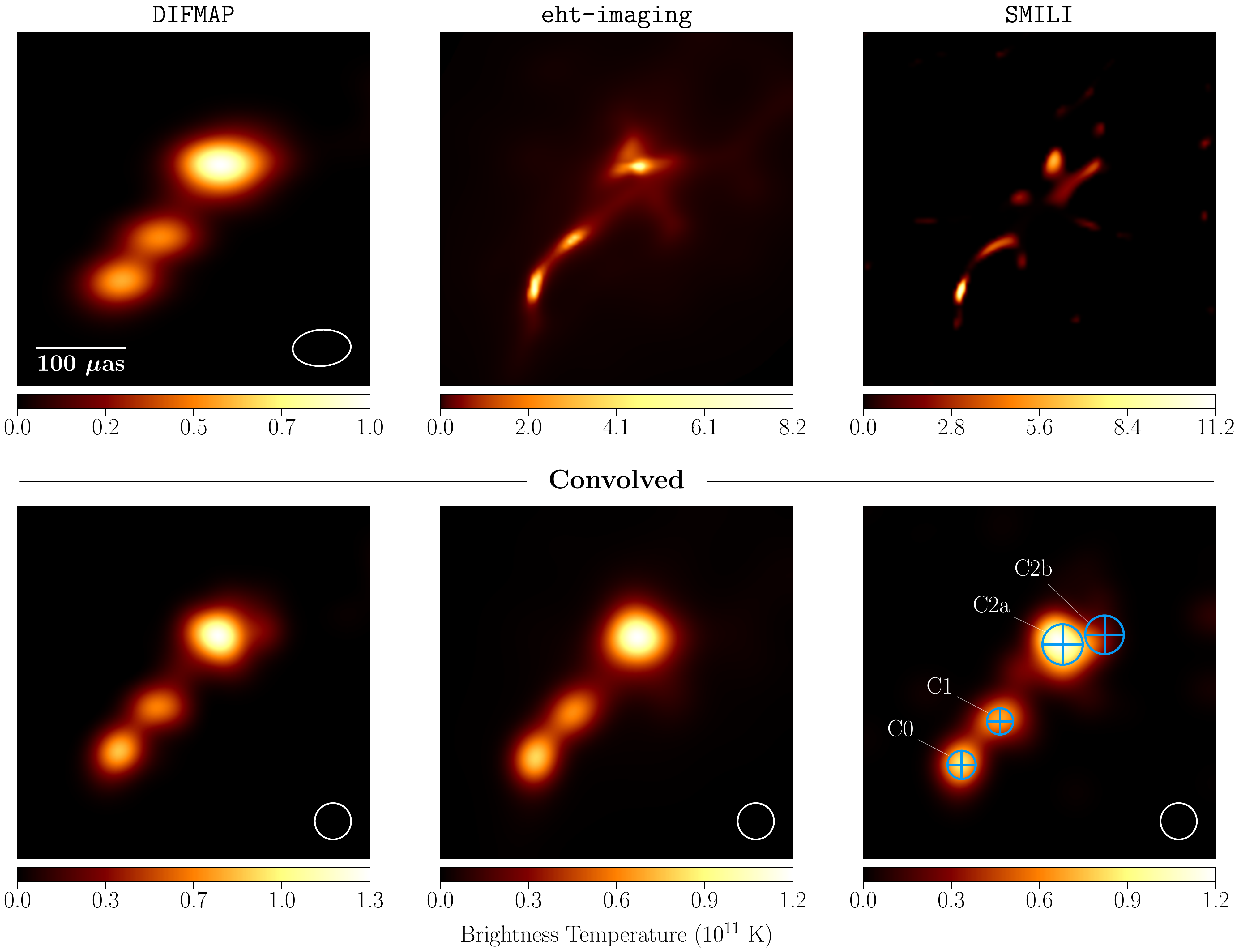}\hfill
\caption{ 
  From left to right: total intensity maps of \oj at 3.5\,mm obtained with GMVA+ALMA observation on April 2, 2017 reconstructed with \difmap, \ehtim, and \smili, respectively. 
  The x and y-axis in each image represent the right ascension and declination axis on the sky, respectively.
  The \difmap image is convolved with the natural-weighted beamsize of the array, which is 64\,{\textmu}as$\times$40\,{\textmu}as at a position angle of $-86$ degrees. 
  For the \difmap, \ehtim, and \smili images, respectively, the reduced $\chi^{2}$ of closure phases is: 1.21, 1.22, 1.19; and that of log closure amplitudes is: 1.18, 1.22, 1.08.
  The second row shows the same images but convolved with a circular beam of 40\,{\textmu}as. 
  The bottom-right panel shows the model-fitted circular Gaussian components overlaid on the convolved SMILI total intensity map. The flux, location, and size of each component are listed in \autoref{tab:modelfit}. 
 }
\label{fig:maps}
\end{figure*}
  
\subsection{Jet Morphology}
\autoref{fig:maps} shows the total intensity maps of \oj obtained with our GMVA+ALMA observations, achieving the highest angular resolution to date of the source at the wavelength of 3.5\,mm. 
The imaging results are consistent across different imaging methods (CLEAN \& RML). Under the nominal resolution, the jet appears to consist of three major features, extending along the southeast to northwest direction. We denote the three features as components C0, C1, and C2, as shown in the bottom right panel of \autoref{fig:maps}.  

Component C0, which lies at the southern end of the jet, is compact and shows the highest brightness temperature~(\autoref{tab:modelfit}). This feature is more likely to be the VLBI core at 3.5\,mm. The component C2 has the highest flux density among the three components. This feature shows complex substructures under the fine resolution of the RML images (\autoref{fig:maps} top-middle \& top-right).  
We see hints of the jet bending and extending towards the western direction downstream of C2. 
This bend is more obvious in the lower frequency maps 
which are more sensitive to the extended lower brightness regions despite the lower angular resolutions \citep[e.g.,][]{galaxies5010012, Jorstad_2017}.
The downstream jet is largely resolved out and not well-constrained in our high-resolution images because of their steep spectra and extended structure. 
Our higher-resolution images reveal for the first time the twisted morphology of the innermost, ultra-compact jet region.  
The first bending occurs between C0 and C1, with the jet axis gradually changing from north to northwest (clockwise). We see also hints for a subsequent bending happening downstream of C1 where the jet axis turns towards the counter-clockwise direction. 

The three-component structure is also consistent with the recent 22\,GHz \textsl{RadioAstron} space-ground VLBI observations of \oj made at a similar resolution \citep[][]{2022ApJ...924..122G}. 
However, a position angle difference of $\sim$\,50$^{\circ}$ of the inner jet can be found when comparing with the \textsl{RadioAstron} image obtained in 2014. Such a difference could be attributed to the variation in the position angle in $\sim$ three years. A detailed analysis of the inner jet position angle variation on a yearly scale and the comparison with theoretical predictions will be presented in a forthcoming paper~(Zhao et al. in preparation).   

\begin{figure}[t]
\centering
\includegraphics[width=1.0\linewidth]{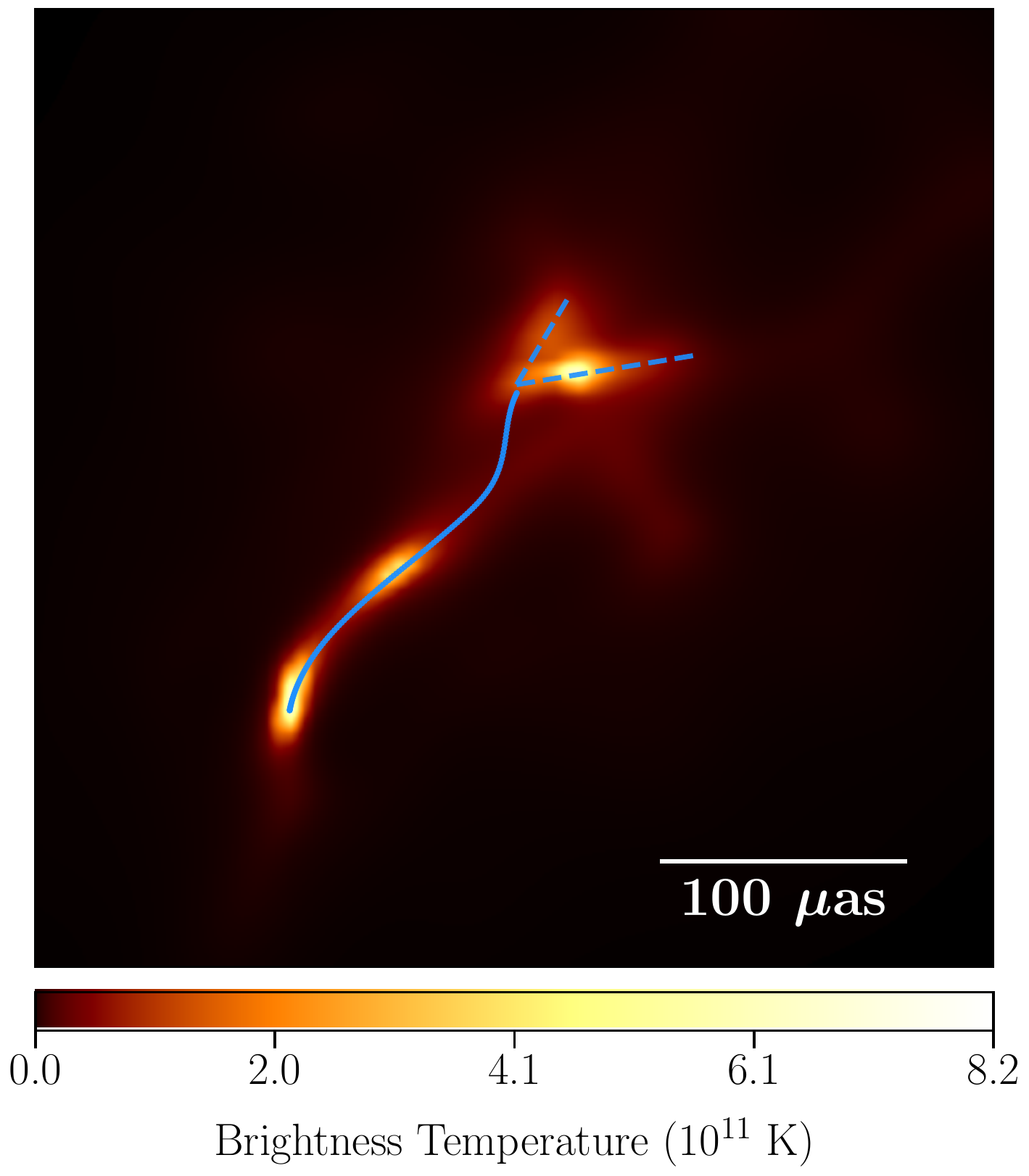}
\caption{
The continuous blue line traces the ridge line of the inner jet of \oj overplotted on the \ehtim reconstructed super resolution image. The dashed blue lines represent the conical structure of the C2 component.
}
\label{fig:ridge}
\end{figure}

In order to quantify the position angle evolution along the jet, we fit the jet ridge-line on the \ehtim map.  
First, we transform the image to polar coordinates centered at the jet origin and slice it transversely. For each slice, we store the flux density peak position and then transform them back to Cartesian coordinates. Thus, we obtain a collection of positions tracing the jet axis between C0 and C2. The results are presented in \autoref{fig:ridge}, where we also show a sketch to trace the conical structure of C2 in the figure.
The jet axis near C0 extends along a position angle of $\sim$\,$-15^{\circ}$, decreases to $\sim$\,$-50^{\circ}$ at C1, and starts to increase again near C2. 
A similar trend can be found also in the \smili image.


\subsection{Brightness Temperatures}
\begin{figure}[t]
\centering
\includegraphics[width=0.999\columnwidth]{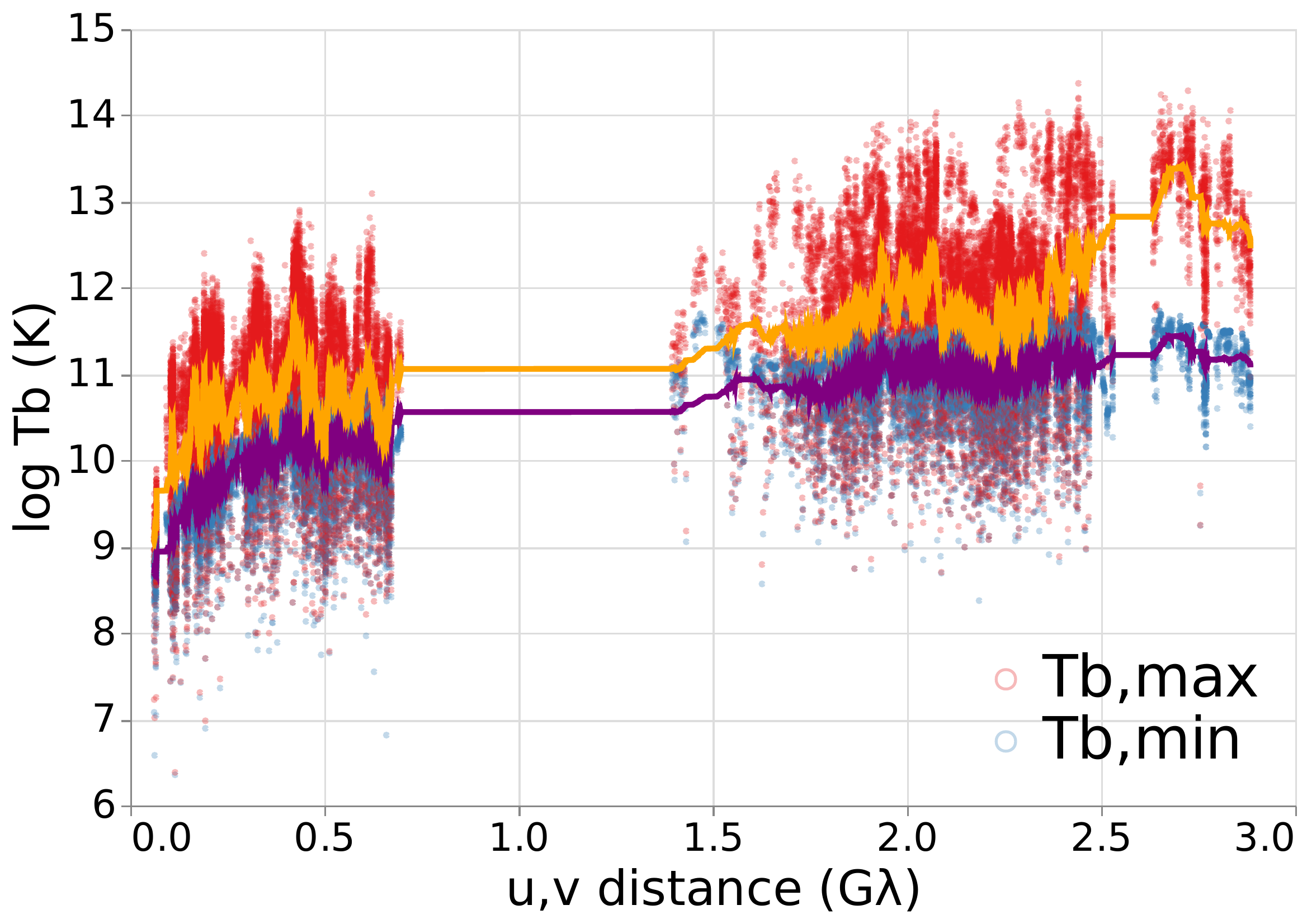} \hfill
\caption
  { 
  Visibility-based brightness temperature estimates of \oj at 86\,GHz using the method described in \citet{2015A&A...574A..84L}. The red and blue dots are the values of $T_\mathrm{b,max}$ and $T_\mathrm{b,min}$, respectively. The orange and purple curves are the rolling mean of the $T_\mathrm{b,max}$ and $T_\mathrm{b,min}$ values. 
  }
\label{fig:tb}
\end{figure}

We investigate the brightness temperature of the \oj jet using two independent approaches: 1) we calculate the observed brightness temperature of each Gaussian component from the model-fitting results using the following equation \citep[e.g.,][]{2002ApJS..141..311T}:
\begin{equation}
   T_\mathrm{b}^\mathrm{obs}=1.22 \times 10^{12} \frac{S}{\theta_{obs}^{2} \nu^2} .
\end{equation}
\noindent
where $S$ is the component flux density in Jy, $\theta_\mathrm{obs}$ is the size of the emitting region in mas, and $\nu$ is the observing frequency in GHz.
2) we calculate the minimum and maximum brightness temperature directly from the visibilities using the method described in \citet{2015A&A...574A..84L}.
The model fitting results, which are listed in \autoref{tab:modelfit}, show the observed brightness temperature of the jet components at 86\,GHz ranges from $10^{10}$ to $10^{11}$\,K. This is in agreement with the values calculated from the visibility amplitudes as shown in \autoref{fig:tb}.  
The brightness temperature values agree quantitatively with the typical values at the same frequency band \citep[e.g.,][]{Lee:2008ir, 2019A&A...622A..92N}.
The 86\,GHz brightness temperatures are about one order of magnitude lower compared to those at 22\,GHz obtained from the \textsl{RadioAstron} results \citep{2022ApJ...924..122G}. 
This can be attributed to differences in intrinsic brightness and opacity between the two frequencies.

We estimate the intrinsic brightness temperature, $T_\mathrm{b}^{\mathrm{int}}$, by \citep[e.g.,][]{2016ApJ...817...96G}: 
\begin{equation}
    T_\mathrm{b}^\mathrm{int}= \left(1+z \right) \delta^{-1} T_\mathrm{b}^\mathrm{obs}
\end{equation}
where $\delta$ stands for the Doppler factor. 
We adopt the value of the latest estimates based on the proper motion of moving components by the VLBA-BU-BLAZAR monitoring program, $\delta = 8.6 \pm 2.8$~\citep{2022arXiv220212290W}.
This gives the intrinsic brightness temperature values 
$T_\mathrm{b,C0}^\mathrm{int}=(3.2 \pm 1.7) \times10^{10}\text{ K, }T_\mathrm{b, C1}^\mathrm{int}=(2.8 \pm 1.4) \times10^{10}\text{ K, }T_\mathrm{b, C2a}^\mathrm{int}=(3.1 \pm 1.4) \times10^{10}\text{ K, and }T_\mathrm{b, C2b}^\mathrm{int}=(0.7 \pm 0.4) \times10^{10}\text{ K}$, for each component, respectively. 
These values fall below the equipartition value of $\sim 5\times 10^{10}$ \citep{1994ApJ...426...51R}, indicating possible magnetic dominance in the innermost jet.
However, this is quite uncertain as the errors in the Doppler factor and brightness temperature values are large.

\subsection{Polarization}
\begin{figure*}[t]
\centering
\includegraphics[width=\linewidth]{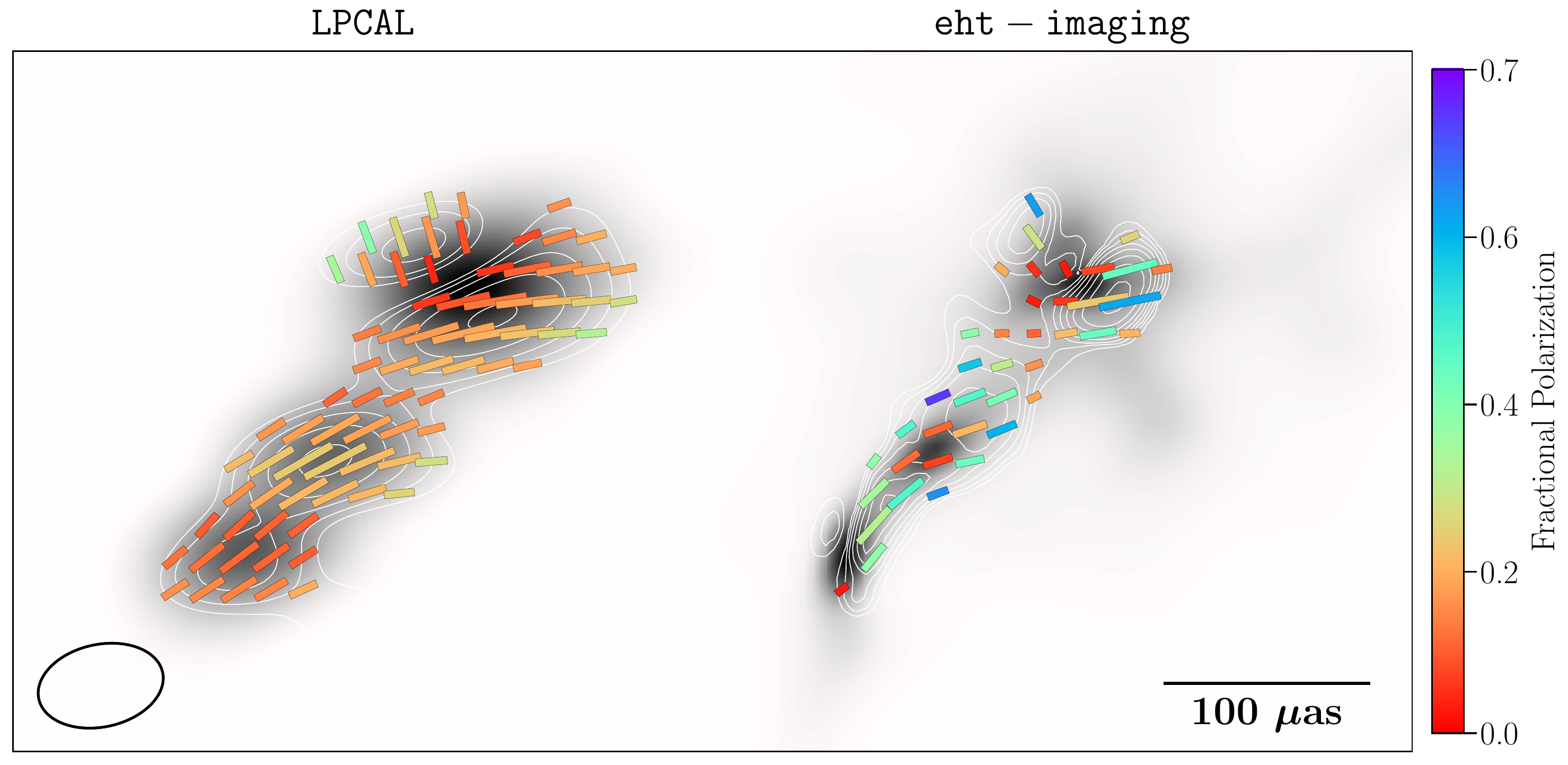}
\caption{Polarized images of \oj produced by \textsc{lpcal} +\textsc{clean} 
method (left) and the RML imaging method \ehtim (right). 
The total intensity image is shown in a grayscale. 
The contours represents the linearly polarized flux density.
The ticks show the orientation of the EVPAs where the lengths indicate the polarization intensity magnitude, and the color represents the fractional polarization. Only the \textsc{lpcal}+\textsc{clean} image is convolved with the beam, shown in the bottom-left.
}
\label{fig:pol_map}
\end{figure*}

We perform polarimetric imaging of the instrumental polarization calibrated data with CLEAN and \ehtim independently. The corresponding images are shown in \autoref{fig:pol_map}. Our images show that the overall degree of polarization of \oj is $\sim$\,8\,\%, which is in quantitative agreement with the ALMA array results of 8.8\,\% presented in~\cite{Goddi21}. The EVPAs extend mostly along the mean jet axis, which suggests that the magnetic field in the jet has a predominant toroidal component. 
Again, the image reconstructed by \ehtim shows fine structure because of the super-resolution that is naturally achieved by the forward modeling method. However, even in the CLEAN image, which is convolved with the nominal beam, we see a remarkable polarimetric structure in the inner jet. The overall structure is consistent between the two images reconstructed independently with different approaches. We notice that the apparent difference in the fractional polarization between the two maps is due to the fact that CLEAN images are convolved with the nominal beam. The overall degree of polarization and the EVPA distributions agree well between the two images. 

Among the several jet components, C0 shows the lowest fractional polarization of $\sim$\,5\% as measured from the \ehtim map. 
This further supports this component as the jet core, which is usually depolarized~\citep[e.g.,][]{2005AJ....130.1389L}. 
C1 exhibits a high level of polarization ($\sim$\,16\%) which indicate the magnetic field is more ordered in this region.  
C2 shows a conspicuous polarimetric structure which can be further divided into two subcomponents with the EVPAs lying perpendicularly to each other. 
The EVPA in the upper subcomponent also lies perpendicularly to the direction along which the brightness extends, while in the bottom subcomponent they lie nearly in parallel.
The degree of polarization is $\sim$\,7\% and $\sim$\,13\% in the upper and lower sub-component, respectively.
These substructures are clearly seen in both the \ehtim and CLEAN images.
    
    
\section{Discussion}
\label{Sec:Discussion}

\subsection{Nature of the C2 Region}
Our GMVA+ALMA observations have revealed a remarkable structure of the inner jet of \oj because of the improved $(u,v)$-coverage and high sensitivity.
In particular, component C2 shows a complex conical structure in both total and linearly polarized intensity, and a bimodal distribution in the EVPAs. 
Previous multi-epoch observations show that this component is nearly stationary~\citep[e.g.,][]{Jorstad_2017, 2017A&A...597A..80H, Lico22}. 
In the following, we discuss the possible nature of this component.

Oblique shocks could result from the jet striking a cloud of interstellar media. Under the precessing jet model, this would naturally happen for some period as the jet sweeps through the ambient material. Since the location of C2 coincides with where the jet bends, the northeastern section of C2 could be interpreted as an oblique shock on one side of the jet. The oblique shock is in a plane making a small angle to the jet boundary on the north-east side. The flow is then bent by the shock toward the west. The magnetic field could get compressed to strengthen the component nearly parallel to the jet. Therefore, the EVPA on the north-east side is roughly perpendicular to the jet. The southwestern section of C2 could then just be the main jet after the bend, with the magnetic field transverse to the jet direction at that point, as usual for a BL Lac object.

Conical shock waves can be formed when there is a pressure imbalance between the jet plasma and the ambient medium. %
The properties of shocks in relativistic jets have been explored by numerical and semi-dynamical simulations.   
\citet{1995ApJ...449L..19G} carried out relativistic hydrodynamics (RHD) simulations of a 
parsec-scale jet surrounded by ambient medium with constant or decreasing pressure. 
The simulations confirmed the existence of stationary components associated with recollimation shocks. 
\citet{Gomez:1997gq} simulated the interaction of standing shocks and relativistically moving perturbations propagating down the stable jet and found that the shock could enhance the emission of the moving feature and the stationary component could be temporarily ``dragged" downstream. 
Further simulations of the interaction between recollimation shocks and traveling shocks are presented in~\citet{2016A&A...588A.101F}, 
based on the observations presented in~\citet{2011A&A...531A..95F, 2013A&A...557A.105F, 2013A&A...551A..32F, 2015A&A...576A..43F}, 
for the particular case of CTA\,102.
Various configurations of the upstream magnetic field components are also included in subsequent numerical simulations~\citep[e.g.,][]{Broderick:2010dy, 2011ApJ...737...42P, 2018ApJ...860..121F}. 
In particular, \citet{2015ApJ...809...38M} studied the kinematically-dominated jets with different magnetic field configurations including axial, toroidal, and helical based on a relativistic magnetohydrodynamics (RMHD) simulation code.
\citet{2018ApJ...860..121F} characterized the properties of recollimation shocks in RMHD simulations of jets at the parsec scale as a function of the dominant type of energy: internal, kinetic, or magnetic. By solving the radiative transfer equations for synchrothron radiation using as input these simulations, they analyzed the total intensity and linear polarization signatures imprinted in the stationary components associated with these shocks. \citet{2021A&A...650A..61F} extended the analysis to RMHD jet models threaded by helical magnetic fields with larger magnetic pitch angles, and explored as well the effect of different non-thermal particle populations on the polarimetric properties of stationary features and the overall observed synchrotron radiation.  

On the other hand, \citet{Cawthorne:1990jo} established a semi-dynamical model assuming only the shock front is emitting and found that conical shock waves could result in polarization angles either parallel or perpendicular to the jet axis. 
This model also considered only random magnetic fields in the upstream jet.
In \citet{Cawthorne:2006kd}, a poloidal magnetic field component was added to the model, and the results can explain well the observed polarization of the knot K1 in 3C\,380. 
Furthermore, \citet{Cawthorne:2013bh} extended this model to include a paired collimating and decollimating shock and the predicted EVPA could successfully describe the observational results of the BL\,Lac object 1803+784.

Comparing our observational results of C2 with the numerical and semi-dynamical studies, we find that the conical shape of the emitting region is quite consistent between our observation and the simulation works. 
Numerical simulations predict a series of stationary shocks along the jet that can be triggered by a pressure imbalance between the jet and the external medium. The reason we find only one conical-shaped component is most likely the adiabatic expansion of the jet. As also shown in \citet{1995ApJ...449L..19G}, with decreasing pressure downstream of the jet, the intensity of the stationary components gradually decreases and the separation between components increases, so the downstream shocks may be too faint and become undetectable at our observing frequency. 
Regarding the polarized emission, the semi-dynamic simulations show different EVPA distributions across the cone. However, the EVPA pattern is more symmetric with respect to the cone axis. 
Numerical simulations also show that the EVPA pattern will depend on the upstream magnetic field configuration and the viewing angle~\citep[e.g.,][]{2015ApJ...809...38M, 2016ApJ...817...96G, 2021A&A...650A..61F}.
\citet{2021A&A...650A..61F} pointed out that jets with a large magnetic pitch angle, i.e., threaded by a helical magnetic field dominated by its toroidal component, can exhibit a bimodal EVPA distribution around recollimation shocks for small viewing angles. This EVPA configuration could imply a sign flip of the Stokes $\mathcal{Q}$ parameter that leads to a EVPA flip, which then results in a dip in the linearly polarized emission, as we observe in the C2 component from the reconstructed polarimetric images.   

Alternative to the standing shock scenario, the observed properties of the C2 component could be a result of geometric effects due to the bending of the jet axis towards the line of sight. 
With a decreasing viewing angle, the enhanced Doppler boosting could amplify the emission in this region and make C2 the brightest component in the inner jet.
If the viewing angle becomes smaller than the jet opening angle, the bimodal distribution of the EVPAs could be produced by the existence of helical magnetic fields in the jet as the direction of the projected magnetic field is different across the component~\citep{2021A&A...650A..61F}. 
This scenario is supported by previous observations which revealed the existence of a bending around C2~\citep[e.g.,][]{Jorstad_2017, 2017A&A...597A..80H, 2022ApJ...924..122G}.
However, it is difficult to explain the conical shape of the emission region with this assumption.

Moreover, by means of multi-epoch GMVA observations, \citet{Lico22} identified a new jet feature in the region of C2, in a quasi-concurrent GMVA observing epoch. The authors argue that the passage of this new jet component through the stationary feature at 0.1\,mas core-separation (i.e., C1) triggered the high energy outburst during 2016-2017~\citep{2017IAUS..324..168K, 2021ApJ...923...51K} including the faint VHE flare detected during February 2017~\citep{2017ATel10051....1M}~\footnote{In fact, the high X-ray flux detected during the Swift MOMO program of \oj triggered the VHE observations, which led to the first VHE detection.} and moved down to the C2 jet region at the time of these observations. In this scenario, the component C2 in our observations could correspond to the blending of the new feature and the standing shock. 
The observed bimodal distribution of the EVPAs could be due to different polarimetric properties of the two components. A similar case was found in the core region of PKS~1510-089 during a $\gamma$-ray flare in 2015~\citep{2019ApJ...877..106P}.

\subsection{Testing the SMBBH model}
\oj is one of the most promising candidates to harbor a SMBBH system at the center. 
In fact, \oj is among the candidates for hosting a nano-Hz gravitational wave emitting SMBBH system~\citep{2021Galax..10....1V}. The binary model has been successful in explaining the periodic light curves and predicting upcoming impact flares, which were confirmed by observations within a few hours~\citep[e.g.,][]{2020ApJ...894L...1L}.   
The direction of the jet axis was also found to be varying with time and this could be also related to the orbital motion of the BHs~\citep[][]{2021MNRAS.503.4400D}.   
Models that do not require a secondary BH to explain the observed variability have also been proposed. For instance, the flux variation could be explained by viewing angle changes and Doppler beaming effects of a precessing jet. The precession could be driven by the Lense-Thirring effect due to the misalignment between the BH spin and the accretion disc~\citep[e.g.,][]{2020MNRAS.499..362C, 2018MNRAS.474L..81L, 2021MNRAS.507..983L, 2018MNRAS.478.3199B}. MHD instabilities (current-driven or Kelvin-Helmholtz) would be also possible to produce helical distorted jet structure~\citep[e.g.,][]{Mizuno:2012ff, Perucho:2012fd, 2019A&A...627A..79V}.

\citet{2021MNRAS.503.4400D} established a model to explain the parsec-scale jet direction variations at different frequencies in which the jet precession is powered by the SMBBH with parameters constrained by optical observations. This model predicts the 86\,GHz jet axis should be $\sim$\,$-37^{\circ}$ around April, 2017 assuming a disc model. The position angles of the inner jet components (e.g., C1, \& C2a) measured in our GMVA+ALMA observation agree well with this prediction (see~\autoref{tab:modelfit}). However, we note that this agreement is partially due to the observing epoch being not far apart from the 86\,GHz GMVA data used to constrain the model. Furthermore, this agreement will not rule out other possible scenarios. For example, the tilted accretion could also result in precession of the inner jet. 
\citet{2018MNRAS.478.3199B} argue that the PA change observed at 15\,GHz can be modeled by a jet precession combined with a nutation of the axis. The precession could be a result of Lense-Thirring  effects and a secondary BH is not always required.
Furthermore, our RML images also revealed a twisted pattern of the innermost jet that resembles a precessing jet in projection. 

Future kinematic studies with multi-epoch GMVA and EHT observations will hopefully provide further insights to distinguish among different theoretical models for the underlying nature of the source. 


\citet{2021MNRAS.503.4400D} also explored the possibility of the existence of a jet from the secondary SMBH based on the SMBBH model. 
With the high sensitivity and improved north-south resolution because of the participation of ALMA, we found no evidence for a secondary jet, even in the \ehtim and \smili images with super resolution.   
There could be several possible reasons for such a non-detection.   %
First, the jet is likely to be short-lived, as commented on in \citet{2021MNRAS.503.4400D}.
Since the projected separation of the two SMBHs in April 2017 is $\sim$\,10\,{\textmu}as~\citep{2018ApJ...866...11D}, the current image resolution is not sufficient to spatially resolve the binary system if there is no extended jet emission from the secondary SMBH. 
The same would apply if the secondary jet extends in a similar direction as the primary jet.
If the secondary jet is present and points in a different direction, the non-detection implies that the brightness temperature of the jet must be lower than $4 \times 10^{9}$\,K, which corresponds to three times the r.m.s. level of the \ehtim map.
We note the dynamic range of our image reconstruction is much higher than the mass ratio of the two BHs. 

We further note that the GMVA+ALMA observations presented in this work are part of a multiwavelength observing campaign of \oj. Close in time observations with the EHT at 230\,GHz (on April 4, \& 9, 2017) and with the \textsl{RadioAstron} space-VLBI mission at 22\,GHz (on March 7, 2017) could provide even higher angular resolutions and probe slightly different regions of the inner jet.  
Together with the observations at X-ray and optical bands~\citep[e.g.,][]{2017IAUS..324..168K, 2020MNRAS.498L..35K, 2021ApJ...923...51K, 2021Univ....7..261K, 2021MNRAS.504.5575K}, we will be able to test or obtain constraints on the physical parameters of the possible jet associated with the secondary SMBH.
 
\section{Summary}
\label{Sec:Summary}
We have carried out GMVA+ALMA observations of \oj on April 2, 2017, which is the first VLBI observation with the phased-ALMA. 
The improved north-south resolution and array sensitivity together with the newly developed RML methods have enabled us to obtain high fidelity, super-resolved images of the \oj jet with unprecedentedly high angular resolution. The convolved RML images also agree with the CLEAN reconstruction. The images have revealed a twisted structure in the innermost region of the jet. 
Our result suggests that the C0 component lying at the southeastern end of the jet is more likely the VLBI core as it is bright, compact, and relatively depolarized. 
The component C2 located at $\sim$\,200\,{\textmu}as northwest of the core shows a conical morphology and complex substructures in polarization. We argue that this component could be an oblique or recollimation shock, or related to a traveling component passing through a stationary feature in the jet.   
We have also carried out the first attempt to search for a jet from the secondary black hole as proposed by \citet{2021MNRAS.503.4400D} based on the SMBBH model.
The non-detection could be due to the small projected separation, the short lifetime, or the difference in the physical conditions of the secondary jet. The EHT and \textsl{RadioAstron} observations carried out in 2017 and later could provide further tests of the SMBBH model.


\section*{acknowledgments}
The work at the IAA-CSIC is supported in part by the Spanish Ministerio de Econom\'{\i}a y Competitividad (grants AYA2016-80889-P, PID2019-108995GB-C21), the Consejer\'{\i}a de Econom\'{\i}a, Conocimiento, Empresas y Universidad of the Junta de Andaluc\'{\i}a (grant P18-FR-1769), the Consejo Superior de Investigaciones Cient\'{\i}ficas (grant 2019AEP112), and the State Agency for Research of the Spanish MCIU through the ``Center of Excellence Severo Ochoa" award to the Instituto de Astrof\'{\i}sica de Andaluc\'{\i}a (SEV-2017-0709).
This publication acknowledges the project M2FINDERS that is funded by the European Research Council (ERC) under the European Union’s Horizon 2020 research and innovation programme (grant agreement No 101018682).
L.L. acknowledges the support of the DGAPA/PAPIIT grants IN112417 and IN112820, the CONACyT-AEM grant 275201 and the CONACyT-CF grant 263356.
TS was supported by the Academy of Finland projects 274477, 284495, 312496, and 315721.
Y.K.\ was supported in the framework of the State project ``Science'' by the Ministry of Science and Higher Education of the Russian Federation under the contract 075-15-2020-778.
R.-S. L is supported by the Max Planck Partner Group of the MPG and the CAS, the Key Program of the National Natural Science Foundation of China (grant No. 11933007), the Key Research Program of Frontier Sciences, CAS (grant No. ZDBS-LY-SLH011), and the Shanghai Pilot Program for Basic Research – Chinese Academy of Science, Shanghai Branch (JCYJ-SHFY-2022-013).

This paper makes use of the following ALMA data: ADS/JAO.ALMA\#2016.1.01116.V. ALMA is a partnership of ESO (representing its member states), NSF (USA), and NINS (Japan), together with NRC (Canada), MOST and ASIAA (Taiwan), and KASI (Republic of Korea), in cooperation with the Republic of Chile. The Joint ALMA Observatory is operated by ESO, AUI/NRAO, and NAOJ. The ALMA data required non-standard processing by the VLBI QA2 team (C. Goddi, I. Mart{\'i}-Vidal, G. B. Crew, H. Rottmann, and H. Messias).
This work is based partially on observations with the 100-m telescope of the MPIfR (Max-Planck-Institut für Radioastronomie) at Effelsberg.
This research has made use of data obtained with the Global Millimeter VLBI Array (GMVA), which consists of telescopes operated by the MPIfR, IRAM, Onsala, Mets\"ahovi, Yebes, the Korean VLBI Network, the Greenland Telescope, the Green Bank Observatory (GBT), and the Very Long Baseline Array (VLBA). The VLBA and the GBT are facilities of the National Science Foundation operated under a cooperative agreement by Associated Universities, Inc. The data were correlated at the correlator of the MPIfR in Bonn, Germany.

\software{%
$\mathcal{AIPS}$
\citep{2003ASSL..285..109G}, 
DiFX \citep{Deller_2007}, 
\texttt{DIFMAP} \citep{1995BAAS...27..903S}, 
\ehtim\ \citep{2018ApJ...857...23C}, 
\texttt{ParselTongue} \citep{2006ASPC..351..497K},  
\smili\ \citep{2017AJ....153..159A}, 
\texttt{Altair} \citep{VanderPlas2018}, 
\texttt{Vega-Lite} \citep{Satyanarayan2017}. 
}
\facility{ALMA, GMVA, VLBA, GBT, EVN, Effelsberg, Mets\"{a}hovi Radio Observatory, Onsala Space Observatory, IRAM-30m RT, Yebes-40m RT}

\bibliography{references}{}

\clearpage
\appendix

\section{Calibration of the Instrumental Polarization} \label{appendix::dterms}
Calibration of the instrumental polarization leakage (also known as the D-terms) is required to obtain reliable polarimetric maps of the target.
Each of the two pipelines that we used to perform polarimetric imaging (see \autoref{Sec:Obs}) has independently implemented this calibration step.

\begin{figure*}[h]
        \centering
        \includegraphics[width=\textwidth]{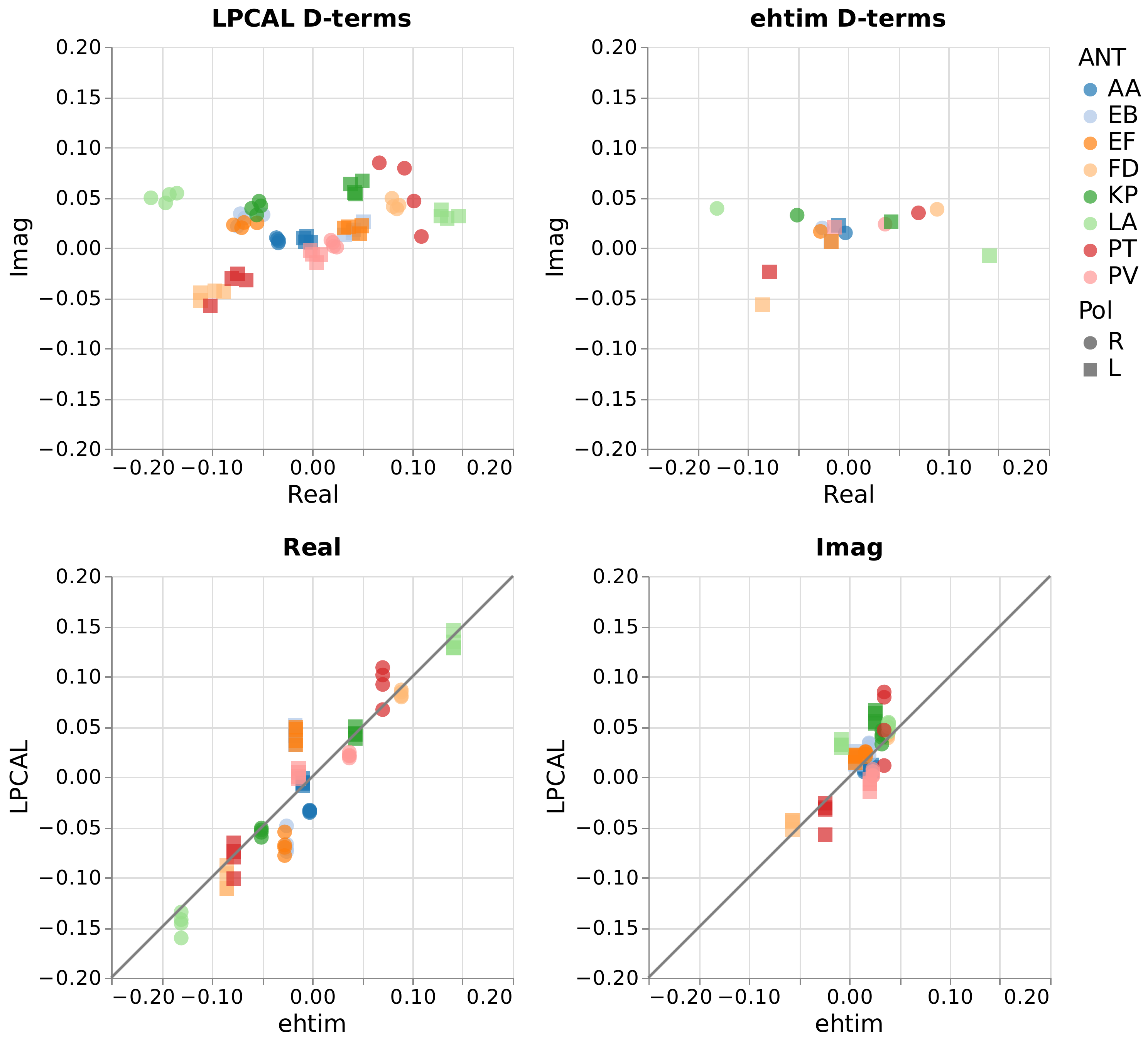}
        \caption{Top: D-term solutions for each station obtained with \textsc{lpcal} (left) and \ehtim (right). \textsc{lpcal} solves for each of the four IFs independently, while in \ehtim the data for different IFs are averaged.
        Bottom: comparison of the Real (left) and Imaginary (right) components of the D-term solutions between \ehtim ($x$-axis) and \textsc{lpcal} ($y$-axis). 
        In all panels, the circular and square symbols represent the data for RCP and LCP, respectively.
        Solutions for different stations are plotted in different colors.
        The station name each abbreviation stands for is as follows: AA: ALMA; EB: Effelsberg (RDBE); EF: Effelsberg (DBBC2); FD: VLBA Fort Davis; KP: VLBA Kitt Peak; LA: VLBA Los Alamos; PT: VLBA Pie Town; PV: IRAM-30m.
        }
        \label{fig:dterms}
   \end{figure*}

The \textsc{lpcal} pipeline loads the self-calibrated visibility data and the CLEAN Stokes I image of \oj produced by \difmap and runs the {$\mathcal{AIPS}$} task \textsc{lpcal} to solve for the D-terms. 
\textsc{lpcal} assumes that the source can be divided into a few sub-components, each with a constant fractional polarization. \textsc{lpcal} solves the D-terms for each IF independently; the results are shown in the top left panel of \autoref{fig:dterms}. We have flagged the stations that only have data for one circular polarization (Yebes-40m) and stations that show low S/N on cross-hands (RL \& LR) polarization data (VLBA NL, \& ON).

On the other hand, the \ehtim pipeline performs the instrumental polarization calibration in parallel with the imaging of the polarimetric data products. 
The pipeline computes the leakage terms by minimizing the difference between the self-calibrated data and the sampled data from the corrupted reconstructions.   %
For details of the polarimetric imaging with \ehtim, refer to \citet{Chael2016}. 
The \ehtim software by default averages the data at different IFs, so we have flagged the stations that show large differences in the D-terms across IFs (VLBA BR, \& OV) in our polarimetric analysis.
The \ehtim results are shown in the top-right panel of \autoref{fig:dterms}.

Despite the different approaches for solving the instrumental leakages, the two pipelines provide very consistent results of D-term estimation, as shown in the bottom panels of \autoref{fig:dterms} which validates our polarization calibration.
The absolute calibration of the EVPA was obtained by comparison with the ALMA observations of \oj at the same frequency performed during the same observation campaign~\citep{Goddi21}. 
    
\end{document}